\documentclass[aps,pre,showpacs,showkeys,superscriptaddress,%
a4paper,%
twocolumn,twoside,floatfix%
]{revtex4}
\usepackage{amsfonts,amsmath}
\usepackage{bm}
\usepackage[dvips]{graphicx}

\newcommand{\beq}{\begin{equation}}
\newcommand{\eeq} {\end{equation}}
\newcommand{\bea}{\begin{eqnarray}}
\newcommand{\eea}{\end{eqnarray}}

\def\vec{\mathbf}

\def\Re{\ensuremath{\mbox{Re}}}

\begin{document}

\title{Turbulence Transition and the Edge of Chaos in Pipe Flow}

\author{Tobias M. Schneider}
\affiliation{Fachbereich Physik, Philipps-Universit\"at Marburg, 
  Renthof 6, D-35032 Marburg, Germany}
\author{Bruno Eckhardt}
\affiliation{Fachbereich Physik, Philipps-Universit\"at Marburg, 
  Renthof 6, D-35032 Marburg, Germany}
\author{James A. Yorke}
\affiliation{University of Maryland, College Park, Maryland, USA}

\begin{abstract}
The linear stability of pipe flow implies that only 
perturbations of sufficient strength will trigger the transition to turbulence. 
In order to determine this threshold in perturbation amplitude we study the \emph{edge of chaos} which separates
perturbations that decay towards the laminar profile and perturbations
that trigger turbulence. Using the lifetime as an indicator and
methods developed in (Skufca et al, Phys. Rev. Lett. {\bf 96}, 174101 (2006))
we show that superimposed on an overall $1/\Re$-scaling predicted and studied
previously there are small, non-monotonic variations reflecting folds
in the edge of chaos. 
By tracing the motion in the edge we find that it is formed by the 
stable manifold of a unique flow field that is dominated by a 
pair of downstream vortices, asymmetrically placed towards the wall. 
The flow field that generates the edge of chaos shows intrinsic chaotic dynamics.
\end{abstract}

\maketitle 
The transition to turbulence in pipe flow has puzzled scientists for more than 
100 years 
because, in contrast to many other flow situations, the laminar profile is 
stable for all 
Reynolds numbers 
and there is no linear instability that could trigger the transition 
\cite{Reynolds1883,Mullin2004,Kerswell2005,Eckhardt2007}. 
Many experiments have therefore focussed on the determination of the `double threshold'
\cite{Boberg1988,Darbyshire1995,Grossmann2000} in Reynolds number and perturbation amplitude that 
has to be crossed 
in order to trigger turbulence. Experimental studies quote a variety of values for 
the Reynolds numbers
that have to be exceeded before sustained turbulence can be observed
\cite{Darbyshire1995,Kerswell2005}. 
They also support a $1/\Re$ scaling for the threshold amplitude \cite{Hof2003}. 
On the theoretical side, numerical studies have shown that for sufficiently high 
Reynolds numbers a variety of 3-d persistent flow structures of travelling
wave type appear in saddle-node bifurcations. For instance,
six symmetrically arranged downstream vortices appear near $\Re=1250$ 
\cite{Faisst2003,Wedin2004,Hof2004}. The critical amplitude has been 
estimated from an analysis of the nonnormal amplification combined 
with an asymptotic analysis of the equations of motion; it gives 
a scaling of the critical amplitude like $1/\Re$ for large Reynolds numbers 
\cite{Grossmann2000,Trefethen2000}. 
For simple models of shear flows, more detailed studies have been possible 
\cite{Eckhardt1999,Moehlis2004,Moehlis2005}. 
In particular, it has been possible to track the 
dynamics at the edge of
chaos which separates initial conditions that decay directly to the laminar profile
and those that swing up to turbulent dynamics \cite{Skufca2006}.
We here apply these ideas to pipe flow, thus contributing to the
elucidation of the key structures in the state space of pipe flow that are responsible
for the transition between laminar and turbulent dynamics. 
The study is part of the dynamical system scenario advocated for the transition 
\cite{Eckhardt2002,Eckhardt2007}, and can be related to similar 
observations in other shear flows,
such as plane Poiseuille flow\cite{Itano2001,Toh2005} and plane Couette flow.

In order to identify the border between laminar and turbulent behavior we 
use the lifetime of perturbations as an indicator 
\cite{Schmiegel1997,Faisst2004}. The lifetime of a perturbation is
defined as the time it takes to come sufficiently close to the laminar profile where 
`sufficiently close' is defined by the requirement that the future evolution is governed 
by the linearized equations of motion, which guarantees that it will 
asymptotically decay.
Within the dynamical system picture of the transition this defines a target region 
around the point in state space that corresponds to the laminar profile. 
For the initial conditions we mimic the experimental
protocol, where the type of perturbation (jets, blowing and suction, 
periodic modulation, etc.) is
predetermined by the setup and where the strength of the perturbation is 
usually controlled with one parameter. We therefore pick a spatial
structure for the velocity field and modify its amplitude, thereby 
scanning the state space along a ray. 
The perturbation we consider here is a pair of vortices as in the
optimally growing modes identified by Zikanov \cite{Zikanov1996},  
modulated in streamwise direction by applying a $z$-dependent tilt 
in order to break translational symmetry:
\beq
\vec{u}_0(r, \varphi,z) = 
\vec{u}_{\text{Zik}} \left(r\,, \varphi + \varphi_0 
\sin\left(\frac{2 \pi}{L} z \right),\,z\right)\;,
\eeq
where $\vec{u}_{\text{Zik}}$ is Zikanov's mode and $L$ is the length of 
the computation domain used in our direct numerical simulation 
In addition to initial conditions generated from Zikanov modes we also used
initial conditions taken from a turbulent run at higher $\mathrm{Re}$,
as in the entry for the March 2006 Gallery of Nonlinear Images 
\cite{Schneider2006},
and obtained convergence to the same dynamics.

For the numerical simulations were used a pseudospectral 
code with about $1.2\,10^5$ degrees of freedom as in \cite{Hof2006}. The code 
was verified by reproducing linear theory, 
turbulent statistics at higher Reynolds numbers 
and the nonlinear evolution of Zikanov modes
\cite{Zikanov1996}. The length of the computational domain is $L=10R$ 
and times are given in units of $R/U_{cl}$, where $R$ is the radius of the pipe and 
$U_{cl}$ the center velocity of the parabolic profile with the same mean flux. 
For the current study we note that a turbulent run up to an observation time 
of 1000 $R/U_{cl}$ takes almost $24$ hours on an $1.5$ GHz IBM Power5 
processor. More than $4000$ integration runs have been carried out.

\begin{figure}
\[
\includegraphics[width=0.380\textwidth]{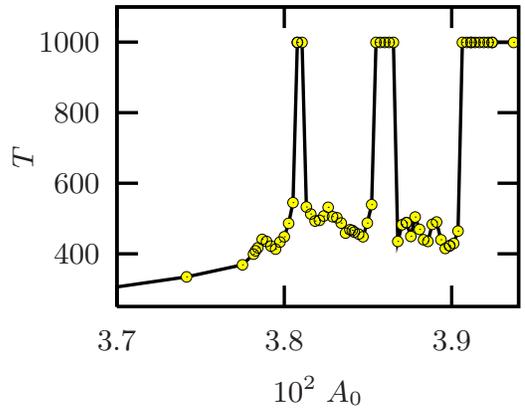}
\]
\\[-1.7em]
\caption[]{(Color online) 
Lifetime $T$ as a function of scaling amplitude $A_0=\sqrt{E_0}$ at a Reynolds 
number of $\Re=3875$. The energy $E_0$ is measured in units of the 
kinetic energy $E_\text{lam}$ of the 
laminar profile generating the same mean downstream velocity. 
\emph{Edge points} clearly separate regions 
where the trajectory turns turbulent and the lifetime reaches 
the numerical cutoff from regions where 
trajectories directly decay.  \\[-1em]
\label{fig:lifetime}}
\end{figure}

A typical variation of the lifetimes with amplitude of the perturbation at fixed 
$\Re$ is shown in Fig.~\ref{fig:lifetime}. In regions with
short lifetimes the flow relaxes quickly to the laminar profile. 
Towards the boundaries of these regions the lifetimes increase quickly and 
reach plateaus at the maximal integration time. 
Magnifications of the plateau regions show chaotic and unpredictable 
variations of lifetimes \cite{Faisst2004}. The cliff-structure in
the lifetimes suggested the name \emph{edge of chaos} for the points at the boundary 
between regions of smooth and of unbounded chaotic lifetime variations, 
respectively \cite{Skufca2006}. 

\begin{figure}
\[
\includegraphics[width=0.380\textwidth]{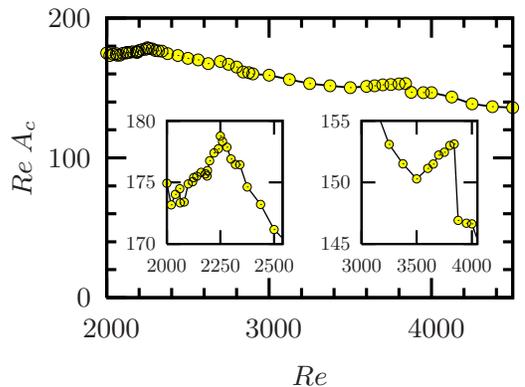}
\]
\\[-1.7em]
\caption[]{(Color online) 
Minimal critical amplitude $A_c = \sqrt{E_c}$ required to trigger 
turbulence times \Re\  as a function of \Re. 
On top of the proposed $1/\Re$ scaling (straight horizontal line) 
several modulations can be observed. The 
insets show magnifications of a `kink' at $\Re\approx 2200$ (left) 
and a `jump' at $\Re \approx 3800$ (right).
\\[-1em]
\label{fig:scaling}}
\end{figure}

The steep increase in lifetimes allows for an accurate tracking of the edge of chaos 
under variations of Reynolds number, see Fig.~\ref{fig:scaling}. 
The edge of chaos lies in an interval in amplitude bounded
by one initial condition which decays towards the laminar profile within the observation 
time of $1000$ units and one that swings up to the turbulent flow. That the upper
trajectory becomes turbulent is also verified by monitoring its energy content. The
widths of the intervals is smaller than the size of the symbols. 
For  
Fig.~\ref{fig:scaling} the amplitude has been multiplied by $\Re$ in order 
to take out the
asymptotic $1/\Re$ scaling \cite{Trefethen2000,Hof2003}. 
In view of the low Reynolds numbers and limited range accessible 
in the numerical study the data are compatible with the expected scaling. 
However, as emphasized by the insets in Fig.~\ref{fig:scaling}, the boundary 
clearly shows kinks, for instance near $\Re=2250$, or 
jumps (near $\Re=3800$) on top of the global  $1/\Re$ scaling. 

\begin{figure}
\[
\includegraphics[width=0.380\textwidth]{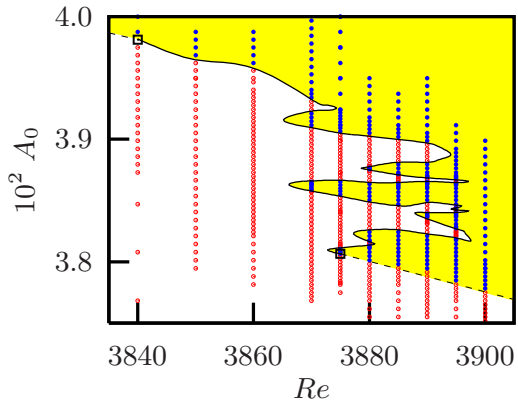}
\]
\\[-1.7em]
\caption[]{(Color online)
Lifetime landscape in the $A_0$-$\Re$ plane. Blue filled circles indicate transition 
to turbulence and red open circles immediate decay. The boundary of the domain 
of initial conditions that connect to the turbulent `state' (shaded) is 
indicated by the folded back line. 
The scaling of the critical amplitude $A_c$ from the inset of Fig.~\ref{fig:scaling} 
is presented as a black dashed line. The jump in $A_c$ from values near $0.0398$ at 
$\Re=3840$ to $0.0380$ at $\Re=3875$ (black open squares) can be directly 
related to the folds in the stability border.  
\\[-1em]
\label{fig:folds}}
\end{figure}

The much finer scan of the lifetimes in the amplitude-Reynolds number plane near
$\Re=3800$ in Fig.~\ref{fig:folds} shows that these structures are due to folds in the
edge of chaos. The two data points to the left and right of the jump in the inset in
Fig.~\ref{fig:scaling} are shown as open squares at $\Re=3840$ and $\Re=3875$. 
The scan at the parameter points indicated reveals regions with long lifetimes,
whose boundary can be connected to form the fingers shown by the continuous 
curve that has been drawn to guide the eye. A kink can then result from an
unresolved small fold. 
This picture is consistent with observations in low-dimensional
models where the variations can be studied with much higher 
resolution \cite{Eckhardt1999,Moehlis2004,Moehlis2005}.
 
\begin{figure}
\[
\includegraphics[width=0.380\textwidth]{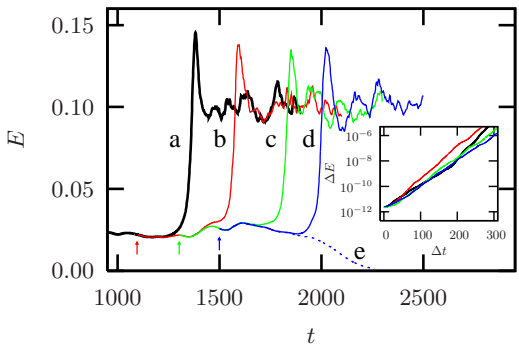}
\]
\\[-1.7em]
\caption[]{(Color online)
Energy traces of trajectories bounding the edge of chaos. The continuous
lines (a-d) show initial conditions that swing up to the turbulent flow 
and belong to the upper end of the interval. For the last control step, 
starting at $t=1500$, also the decaying trajectory (e) from the lower end 
of the interval is presented as a dotted line. 
Control steps 
corresponding to lines (b-d) are indicated by arrows. Trajectory (a) starts at $t=900$.
The inset shows the energy norm of the difference between the two 
bounding trajectories: the uniform expansion for all segments shows that 
they all belong to the same invariant state. 
In absolute values the energy of the difference
increase from initially $10^{-12}$ to at most $10^{-8}$ during one iteration step.  
\label{fig:approach}}
\end{figure}

The monitoring of lifetimes can also be used to track the dynamics 
in the edge of chaos. The studies in \cite{Skufca2006} suggest that all 
edge points at a given Reynolds number are connected and lie on the stable 
manifold of an invariant object that resides in state space between the laminar
flow and the turbulent dynamics. To find this invariant object, we repeat the
steps from \cite{Skufca2006}: first, find the amplitude for two initial conditions 
on either side of the edge of chaos, i.e. one which decays towards the 
laminar profile and one that swings up to the turbulent flow. These two 
trajectories shadow one that stays in the edge for all times.
After about 200 time units we refine the approximation and 
determine a new pair of trajectories close to the edge. 
The pair of initial conditions is found along the ray connecting 
the approximated state from the previous step 
(the one escaping to the turbulence) and the laminar profile.
Four refinements and the exponential separation in energy within a pair 
are shown in Fig.~\ref{fig:approach}.
The inset shows that we keep their separation below $10^{-8}$ in energy.
As long as the direction in state space defined by the amplitude scaling is not 
tangent to the edge manifold this technique can be used to obtain 
arbitrarily long traces of
a state that neither decays nor swings up to the turbulence: 
this trajectory lives in the edge of chaos and should approach 
the invariant state embedded in the edge of chaos the \emph{edge state}.

The edge state is a relative attractor: it is attracting
for initial conditions confined to the edge of chaos, but repelling 
perpendicular to it. In its simplest form the invariant state is a 
hyperbolic fixed point and the edge is its stable manifold.
Numerical studies of the low-dimensional shear flow models 
and of chaotic maps show that the edge state
can also be a periodic orbit or a chaotic relative attractor.

\begin{figure}
\[
\includegraphics[width=0.380\textwidth]{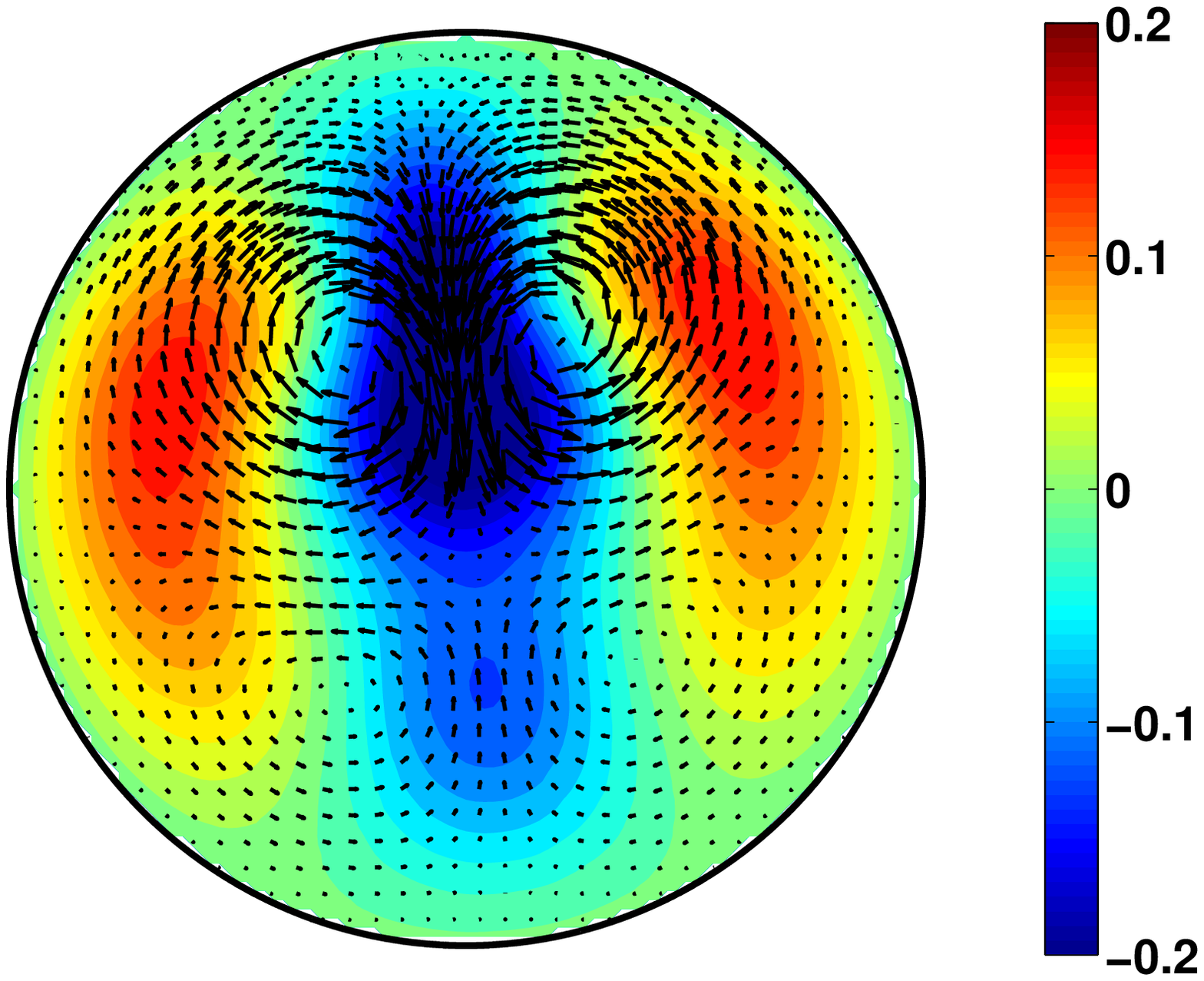}
\]
\[
\includegraphics[width=0.5\textwidth]{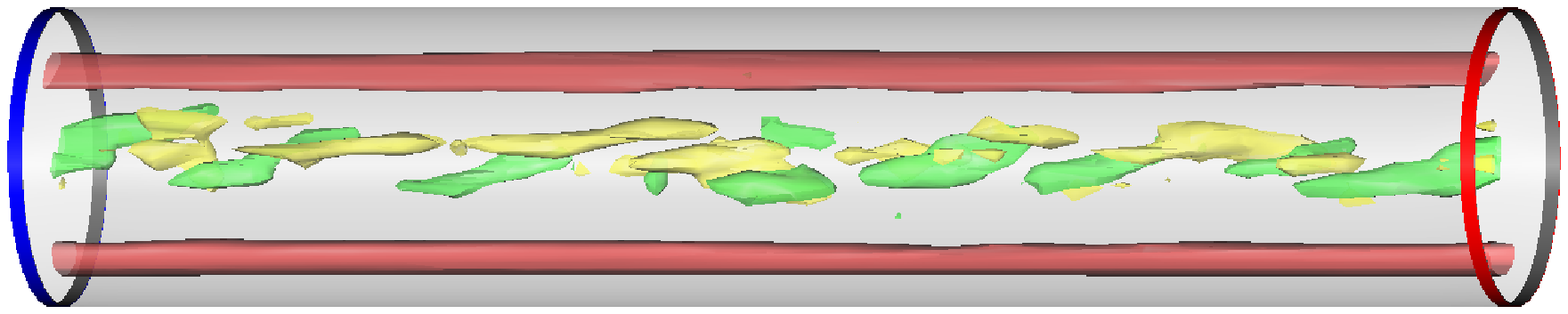}
\]
\\[-1.7em]
\caption[]{(Color online)
The edge state at $\Re=2875$. 
Top: Time-averaged cross section perpendicular to the pipe axis.  
The downstream velocity relative to the parabolic 
laminar profile is shown in color ranging form red (fast) to blue (slow). 
In-plane velocity components are indicated by vectors.    
\label{fig:edgestate}
%
Bottom: The instantaneous flow field along the axis.
Isosurfaces of the downstream velocity (red)
indicate the position of the high-speed streaks. Isosurfaces of the 
downstream vorticity (positive in yellow and negative in green) highlight 
vortical structures in the center region. The fluid flows from left to right.  
\label{fig:sideview}}
\end{figure}

The continuing modulations in the energy trace in Fig.~\ref{fig:approach} 
indicate that the dynamics constrained within the edge of chaos does 
not relax to a stationary or simply periodic state, even when 
followed up to observation times of more than $2000$.
Fig.~\ref{fig:edgestate} shows a cross section of the flow field, averaged 
over a time interval of $200$ to highlight large scale features. 
The global structure of the flow field is simple and dominated by two 
high-speed streaks and a corresponding pair of strong counter-rotating 
vortices which are located off-center. 
It shows no discrete rotational symmetry like the travelling waves studied 
in \cite{Faisst2003,Wedin2004} or the optimal amplification mode 
in \cite{Zikanov1996}. 

The side view of an instantaneous snapshot of the edge state in 
Fig. \ref{fig:sideview} shows no periodicity in the isosurfaces 
of the downstream vorticity, again supporting 
the conclusion that the dynamics in the edge does not settle down to a
simple periodic or quasiperiodic travelling wave. The small scale modulations 
persist and reflect an intrinsically chaotic dynamics of the vortical 
structures in the center region.

Pipe flow in a domain that is periodically continued in the axial direction 
has two continuous symmetries of azimuthal and axial translations. 
Shifted and rotated initial conditions will then give
shifted and rotated edge states. 
Except for these two continuous symmetries, the same edge state is 
obtained for different initial
conditions.
It is intriguing that independent of the original velocity
field the same relative attractor is reached and that this 
relative attractor is dominated by two downstream vortices.

Upon variation of the Reynolds number over the range $2160\ldots 4000$ studied here,
the overall appearance of the edge state does not change much. 
We did not detect any bifurcations or transitions between flow 
topologies, as in the low dimensional
model \cite{Skufca2006}. 
Energetically, the edge state is clearly separated from both the laminar state 
(here the energy of the disturbance vanishes exactly) and also from the 
turbulent state (cf. Fig. \ref{fig:approach}).  
However, in view of the transient nature of the turbulent state \cite{Hof2006}, 
the stable manifold of the laminar profile and the stable manifold of the
edge state have to intermingle tightly in the region with turbulent dynamics. 

The form and topology of the edge state suggests that it can be induced 
experimentally by removing fluid 
at one point near the wall and by injecting fluid at two points to the left 
and right of the removal point.
This perturbation will then reach into the fluid, forming a pair of vortices 
not dissimilar to the
optimally amplifying ones of Zikanov \cite{Zikanov1996}. The vortices will 
then draw energy from the base flow and
induce high- and low-speed streaks in the downstream velocity. As in the 
case of the self-sustained
cycle for near wall turbulence \cite{Hamilton1995,Waleffe1995}, 
one can then anticipate a shear flow instability of the
streak arrangement. A direct visualization of this instability is 
difficult because of the
chaotic dynamics of the edge state, but the exponential growth in 
energy supports the assumption
of an instability. Further evolution of the velocity fields then shows that once the
energy level of the fully developed turbulence is reached, many more 
vortices appear, and the velocity
field shows more signatures of the symmetric vortex arrangements 
\cite{Hof2004,Schneider2007,Kerswell2007}. 
It is satisfying to see that the concepts developed in dynamical system 
theory and verified on
low-dimensional models can be transferred so directly to the full, 
spatially extended systems and that 
they continue to provide insights into dynamics of this old and puzzling problem.

We thank J.~Vollmer and J.~Westerweel for helpful
discussions, \emph{Hessisches Hochleistungsrechenzentrum} in
Darmstadt for computing time, and the Deutsche Forschungsgemeinschaft for support.

\end{document}